\documentclass[aoas, preprint]{imsart}

\RequirePackage{amsthm,amsmath,amsfonts,amssymb}
\RequirePackage[authoryear]{natbib}
\RequirePackage[colorlinks,citecolor=blue,urlcolor=blue]{hyperref}
\RequirePackage{graphicx}

\usepackage{subfiles}

\startlocaldefs
\theoremstyle{plain}

\newtheorem{theorem}{Theorem}

\newtheorem{prop}{Proposition}
\theoremstyle{definition}
\newtheorem{definition}[theorem]{Definition}

\newtheorem*{rem}{Remark}
\usepackage{algpseudocode}
\usepackage{mathtools}
\usepackage{tkz-graph}
\usepackage{algorithm}

\newcommand   \Ocal  {\mathcal{O}}
\newcommand   \Scal  {\mathcal{S}}
\newcommand \Pn {\mathcal{P}_n}
\newcommand{\defeq}{\vcentcolon=}

\endlocaldefs

\begin{document}

\begin{frontmatter}
\newcommand{\letitre}{Set-valued data analysis for interlaboratory comparisons}
\title{\letitre\thanksref{T1}}
%\title{A sample article title with some additional note\thanksref{t1}}
\runtitle{\letitre}
\thankstext{T1}{This article presents results for a subset of the data
gathered during the comparison.
These results by no means replace those of the official report on the results of the comparison.}

\begin{aug}
%%%%%%%%%%%%%%%%%%%%%%%%%%%%%%%%%%%%%%%%%%%%%%%
%% Only one address is permitted per author. %%
%% Only division, organization and e-mail is %%
%% included in the address.                  %%
%% Additional information can be included in %%
\author[A]{\fnms{Sébastien J.}~\snm{Petit}\ead[label=e1]{sebastien.petit@lne.fr}\orcid{0000-0001-7949-3650}},
\author[A]{\fnms{Sébastien}~\snm{Marmin}\ead[label=e2]{sebastien.marmin@lne.fr}}
\and
\author[A]{\fnms{Nicolas}~\snm{Fischer}\ead[label=e3]{nicolas.fischer@lne.fr}}
%%%%%%%%%%%%%%%%%%%%%%%%%%%%%%%%%%%%%%%%%%%%%%
%% Addresses                                %%
%%%%%%%%%%%%%%%%%%%%%%%%%%%%%%%%%%%%%%%%%%%%%%
\address[A]{Department of Data Science and Uncertainty,
Laboratoire national de métrologie\printead[presep={ ,\ }]{e1,e2,e3}}

\end{aug}

\begin{abstract}
This article introduces tools to
analyze set-valued data statistically.
The tools were initially developed to analyze results from an interlaboratory comparison
made by the Electromagnetic Compatibility Working Group of Eurolab France,
where the goal was to select a consensual set of injection points on an electrical device. 
Families based on the Hamming-distance from a consensus set are introduced
and Fisher's noncentral hypergeometric distribution is proposed to model
the number of deviations.
A Bayesian approach is used and
two types of techniques are proposed for the inference.
Hierarchical models are also considered to quantify a possible within-laboratory effect.
\end{abstract}


\end{frontmatter}
%%%%%%%%%%%%%%%%%%%%%%%%%%%%%%%%%%%%%%%%%%%%%%
%% Please use \tableofcontents for articles %%
%% with 50 pages and more                   %%
%%%%%%%%%%%%%%%%%%%%%%%%%%%%%%%%%%%%%%%%%%%%%%
%\tableofcontents

\section{Introduction}

Interlaboratory comparisons (ILC) are required in countless scientific and technical fields
to monitor the performance of laboratory facilities
or to test the relevance of a measurement procedure.
In addition to the financial investment,
organizing such studies is often complex and requires
a significant effort in terms of time and logistics from the organizers.

Standard statistical treatments for evaluating ILC data are very focused on the classic case of measurements resulting in a scalar number. Typically in this case, after simple data processing, a Gaussian assumption is used to evaluate statistically the measurement method or the performance of each participant. To date, to our knowledge, no method exists for set-valued data, where each ``measurement'' result is a subset of a finite set. Certainly, recommendations exist for categorical data, whether ordered or not, but no suitable guidance take into account the subset nature that a measurement may have (e.g., the ISO-13528 standard mentions only expert judgments for scoring categorical data with no natural order), and do not really establish a statistical framework as is the case for real-valued data.
More generally, the authors of this manuscript have not found any suitable methodologies in the statistical literature (see Section~\ref{sec:review} for a review of the literature).

One of the services provided by the French metrology institute, the LNE, is precisely to handle interlaboratory comparisons.
A case of set-valued data was requested in 2023 by the Electromagnetic Compatibility (EMC) Working Group of Eurolab France, an association of certified laboratories.
The benefits of this kind of comparison is twofold for the laboratories.
First, in some cases, this makes it possible to compare decisions that are closer to those made by the operators of these laboratories in real-life situations. 
Secondly, the organization of the comparison can be greatly facilitated. For example, for the EMC comparison, participants were given a photo and a spreadsheet to fill in. 
Whereas, usually, a test material circulates, sometimes for several years, between the participants.

\begin{figure}
\centering
\includegraphics[scale=0.57]{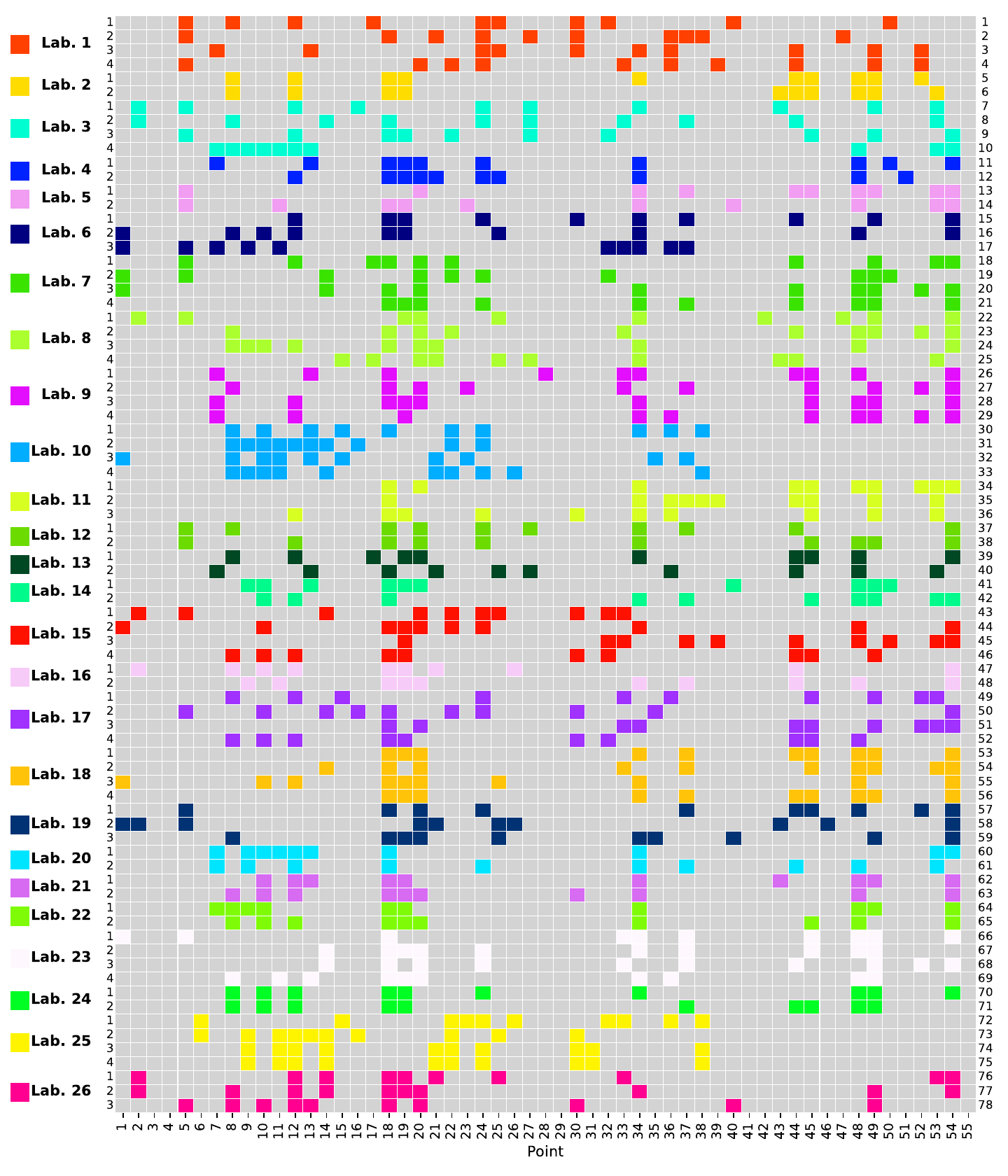}
\caption{Choices of~$78$ operators from~$26$ laboratories.
The~$x$-axis represents the elements of~$\Ocal$ and each row represents the choices made by one operator.
Colors are used to identify laboratories.
}
\label{fig:mozaique_cil_cem}
\end{figure}

The goal of the study was to compare and analyze the results of different laboratories when selecting $10$ electrostatic discharge injection points on an electrical device. 
In Figure~\ref{fig:mozaique_cil_cem}, each selection made by one operator is represented in a row, as a subset of size $10$ of a set of $55$ items.

The task of comparing subset selections is, of course, not limited to this field. A newly available method for rigorously comparing set-valued data could, across various domains, motivate to process data that have remained unexploited or generate numerous ideas for comparisons, thereby ultimately improving the robustness of practices.

In this work, we present the Bayesian model developed as a solution for Eurolab. 
It is able to find  collective set-valued consensus emerging from individual propositions of same cardinal (a generalization to subsets of different sizes could also be possible). As we aim at estimating a central set of a distribution of set, we rely on probability distributions based on the Hamming-distance.
We derive a well-defined and interpretable uncertainty about the most likely consensus set from which we evaluate individual deviations in terms of Bayesian $p$-values.
Moreover, a hierarchical model can be derived for 
quantifying a possible within-laboratory effect.

After a brief literature review in Section~\ref{sec:review}, 
Section~\ref{sec:setup} establishes formal notations using an example. A new statistical model is then developed in Section~\ref{sec:model}. Section~\ref{sec:inference} details the inference methodologies, including MCMC sampling strategies for parameter estimation. The real-life case study on electromagnetic testing is presented in Section~\ref{sec:application}.
Section~\ref{sec:conclusions} presents our conclusions and perspectives for future work.
%Section~\ref{sec:conclusions} discuss implications for standardization. 

%Limitations?

\section{Literature review}
\label{sec:review}

Qualitative comparative analyses \citep[see][]{schneider2012set} use sets for data analysis.
However, in contrast to our task, the goal is to explain known labels by searching for logical relationships between features.

On the other hand, \citet{Fey2014CollectiveCO} studies voting procedures theoretically
in the unranked multiple-choice setting, which is the one of
the present interlaboratory comparison case study.
While the results provide a interesting insight on voting procedures,
the scope does not involve the calculation of the central set in a statistical framework.
%Similarly, while statistical sampling theory \citep[see, e.g.,][]{thompson2012sampling} provides valuable tools,
%it does not directly apply to the present problem, tackling sampling approaches for population surveys.

Fuzzy sets \citep[see, e.g.,][]{zadeh1965fuzzy} are sets whose elements can have different levels of membership.
While fuzzy sets were used to compute median---or centroidal---values for numerical sets \citep{beliakov2011aggregation, friedman1997most},
we could not find a fuzzy logic-based approach for computing a consensual set from candidate sets of same size.

In many works (see, e.g., \citealp{dawid1979maximum}, or more recently \citealp{raykar2010learning}), where a consensual output is derived from different expert agent,  the focus is given on scalar or boolean outputs, not sets (even if set membership could be encoded with boolean outputs, the set size constraint is not enforced with these methods). 

In a recent article, \cite{fellmann2024kernel} propose a kernel for
measuring the importance of inputs of a set-valued model with a
Hilbert-Schmidt Independence Criterion \citep{gretton2005kernel}.
However, this approach does not make it possible to define a consensus set
from data and assess statistically if deviations are significant.

Finally, the possibility to account for random effect (like the assumption that candidate subsets  can be affected similarly if they are originated from the same organization), was  not found in the literature, although being a very common hypothesis in interlaboratory comparison.

\section{Notations and introducing example}\label{sec:setup}

Let $\Ocal$ be a finite set of size~$\# \Ocal = M$ and let $n < M$, $N = M - n$,
and $\mathcal{P}_n$ be the set of all subsets of~$\Ocal$ with size~$n$.
For an~$A \in \Pn$,
the notation~$A^c = \Ocal \setminus A$ will also be used.

Consider some data $X_1, \dots, X_p \in \Pn$.
For instance, suppose that~$\Ocal$ is a set of zones on an industrial part and
some operators are asked to select a subset of~$n$ zones from~$\Ocal$ to be tested.
Table~\ref{tab:example} shows an example of resulting data set.
\begin{table}
\caption{An example of a data set of elements from~$\Pn$, for~$\Ocal = \{1, 2, 3, 4, 5, 6, 7, 8, 9, 10\}$.}
\label{tab:example}
\centering
\begin{tabular}{|c || c|c|c|}
\hline
$X_1$ & $1$ & $2$ & $3$ \\
\hline
$X_2$ & $1$ & $2$ & $3$ \\
\hline
$X_3$ & $1$ & $2$ & $3$ \\
\hline
$X_4$ & $1$ & $2$ & $7$ \\
\hline
$X_5$ & $1$ & $2$ & $3$ \\
\hline
$X_6$ & $1$ & $2$ & $3$ \\
\hline
$X_7$ & $1$ & $2$ & $4$ \\
\hline
$X_8$ & $1$ & $2$ & $3$ \\
\hline
$X_9$ & $2$ & $3$ & $8$ \\
\hline
$X_{10}$ & $1$ & $2$ & $3$ \\
\hline
$X_{11}$ & $1$ & $2$ & $3$ \\
\hline
$X_{12}$ & $5$ & $6$ & $8$ \\
\hline
\end{tabular}
\end{table}

The aim is to judge the ability of operators to identify these positions.
Following the approach used for continuous data, we propose
to analyze the sample statistically to
assess its homogeneity and detect atypical responses.
We assume that no features can be used to establish similarities between the elements of~$\Ocal$.
In this case, it seems reasonable to consider only the numbers of common elements between the samples.

\begin{figure}
\centering
\input{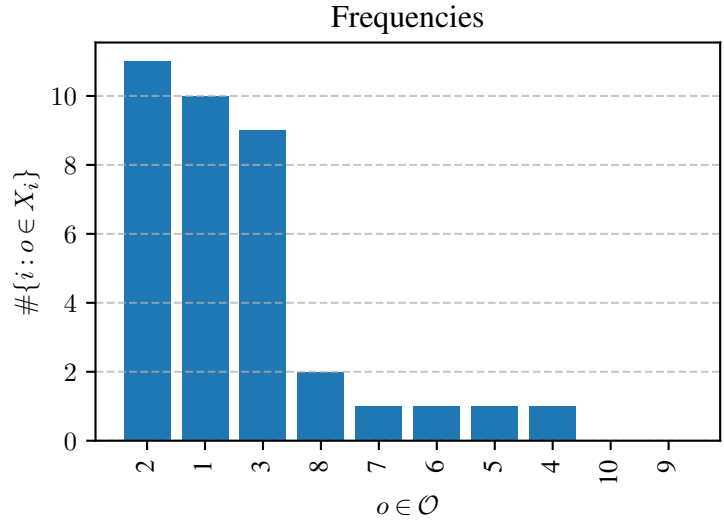}
\caption{The number of~$X_i$s containing each element of~$\Ocal$.}
\label{fig:hist_example}
\end{figure}

Specifically, a natural approach would be to consider the set~$A = \{1, 2, 3\}$
of the three elements that appear most frequently in the data, 
as shown in Figure~\ref{fig:hist_example}.
Considering~$A$ as a consensus value, observe that~$X_{12}$ has no
intersection with~$A$, whereas all the other eleven~$X_i$s have at most
one element outside~$A$.
Are these numbers significant enough to conclude that~$X_{12}$ is an outlier?
What if there was more spread in the data
so that~$A$ could not be unequivocally defined?

The goal of this article is to introduce statistical models to detect outliers
in set-valued data~sets. In particular, significance thresholds are proposed
for the above elementary analysis.
The proposed models are inspired by the usual Gaussian model used for analyzing continuous data.

\section{A statistical model for set-valued data}
\label{sec:model}
\subsection{Hamming-distance probability distributions over subsets}

We introduce for subsets a notion of probability distribution defined from a central set.
Beforehand,
observe that given~$A \in \Pn$ and $0 \leq k \leq n$,
the number of subsets $X \in \Pn$ such that~$ \# \left( X \cap A^c \right) = k$ is~$\binom{n}{k} \binom{N}{k}$.
We will write~$E$
for the set of probability mass functions over~$\{0, \dots, n\}$.
\begin{definition}\label{def:hamming_dist}
For~$A \in \Pn$ and $e \in E$, define the Hamming-distance probability distribution
\begin{equation*}
Q_{A, e} \left( X \right) = \frac{e(k)}{\binom{n}{k} \binom{N}{k}},
\quad
X \in \Pn,
\quad
k = \# \left( X \cap A^c \right).
\end{equation*}
\end{definition}
Under~$Q_{A, e}$, the discrete variable~$k = \# \left( X \cap A^c \right) \in \{0, \dots, n\}$ is \hbox{$e$-distributed}
and, given~$k$,
the subset~$X$ follows a uniform distribution on the
subsets having~$k$ elements outside~$A$.

The following definition formalizes a desirable property of~$Q_{A, e}$.
\begin{definition}\label{def:gaussian_prop}
For~$A \in \Pn$ and $e \in E$,
call~$Q_{A, e}$ Hamming non-increasing if it holds that
\begin{equation*}
Q_{A, e} \left( X \right) \geq Q_{A, e} \left( Y \right)
\end{equation*}
for all $X, Y \in \Pn$ such that $\# \left( X \cap A^c \right) \leq \# \left( Y \cap A^c \right)$.
The probability distribution is called Hamming decreasing when the inequality is strict for~$\# \left( X \cap A^c \right) < \# \left( Y \cap A^c \right)$.
\end{definition}
In particular, this definition ensures that~$A$ is a mode of~$Q_{A, e}$.
It provides an analogy with unimodal probability density functions on~$\mathbb{R}^d$,
which are non-increasing functions of the distance to the mode, such as the Gaussian distribution.
Interpreting the results will be easier for probability distributions that verify this definition.
Section~\ref{sec:binomial} discusses the Hamming non-increasing property when~$e \in E$
is a binomial distribution and Section~\ref{sec:fisher_example} introduces
a family of elements of~$E$ with an even stronger interpretability property.

\subsection{Bayesian statistical model on~$\Pn$}

Let $\underline{X} = \left(X_1, \dots, X_p \right) \in \Pn^p$ be $p$ subsets of~$\Ocal$ of size~$n$.
Hamming-distance probability distributions make it possible to
infer a center for~$\underline{X}$.
To pursue the analogy with Gaussian models for continuous data,
we also aim to infer a kind of dispersion.
According to the model, the number of elements outside the center is distributed
according to~$e \in E$.
To facilitate estimation,
we consider a smaller parameter space by using a parametric
mapping~$e \colon u \in I \to e_u \in E$
and we will write~$(P_{A, u})$ for the family
$$
\left( P_{A, u} \defeq  Q_{A, e_u}, \quad \mathrm{for} \ \ A \in \Pn \ \ \mathrm{and} \ \ u \in I\right)
$$
of probability distributions on~$\Pn$
parametrized by the center~$A \in \Pn$ and the dispersion parameter~$u \in I$.
In the following,~$(P_{A, u})$ will be referred to as a family of Hamming-based probability distributions
and the mapping~$e$ will be clearly specified when necessary.
Typically, the set~$I$ will be an interval of~$\mathbb{R}$
and~$e$ will put all its mass at zero for an endpoint of~$I$.

To infer~$A$ and~$u$,
assume the~$X_i$s are independent and identically distributed from $P_{A, u}$ and write
$$
\mathcal{L} \left( \underline{X}; \, A, \, u \right)
= \sum_{i=1}^p \ln(P_{A, u} (X_i))
$$
for the log-likelihood function.
If~$(P_{A, u})$ is a family of Hamming non-increasing probability distributions,
then an~$A_1 \in \Pn$ is more likely than an other~$A_2 \in \Pn$ if~$A_1$
has uniformly more elements in common with the~$X_i$s. 
For instance, this ensures that~$\{1, 2, 3\}$ is more likely than~$\{1, 2, 10\}$
given the data from Table~\ref{tab:example}.
However, this is not enough to ensure that~$\{1, 2, 3\}$ is more likely than~$\{1, 2, 7\}$,
even if~$3$ is chosen much more often than~$7$. 
Section~\ref{sec:fisher_example} proposes a family of Hamming non-increasing probability distributions
that guarantee that sets whose elements are chosen more frequently are more likely.

As mentioned in Section~\ref{sec:setup}, there may be situations where the choice of~$A$ is subject to uncertainty,
for example, when more than~$n$ objects are more often chosen with similar frequencies.
Consequently, we choose a Bayesian approach to take into account a possible uncertainty on the choice of~$A$.
Without further information, we choose a uniform a prior on~$\Pn$ for the parameter~$A$ and an independent prior for~$u$.

\subsection{Taking a possible within-laboratory effect into account}\label{sec:lab_effect}

Data are sometimes available in the form of selections of points made by groups of operators belonging to different laboratories.
In this case, we will denote the number of laboratories by~$L$
and write
$$
\underline{X}_i = (X_i^{(1)}, \dots, X_i^{(p_i)}) \in \Pn^{p_i},
\quad \mathrm{for} \, 1 \leq i \leq L,
\quad \mathrm{and} \quad \underline{\mathbb{X}} = \left(\underline{X}_1, \dots, \underline{X}_L \right).
$$
We propose models that take into account the fact that operators belong to different laboratories.
One of the objectives is to be able to make statistical statements about biases
that may exist within laboratories when selecting points.
Following the vocabulary used for the analysis of interlaboratory comparisons with continuous data,
if such biases are statistically highlighted, then one says that a ``within laboratory effect'' exists.

\begin{figure}
\centering
\begin{tikzpicture}
    \node[shape=circle,draw=black] (topA) at (0, 0){$A$};
    
    \node[shape=circle,draw=black] (topp) at (0, -1.5){$u$};
    
    \node[shape=circle,draw=black] (A1) at (-4, -3) {$A_1$};
    \node[shape=circle,draw=black] (AL) at (4, -3) {$A_L$};
    
    \node[shape=circle,draw=black] (p1) at (-4, -4.5) {$u_1$};
    \node[] at (0, -4.5) {$\cdots$};
    \node[shape=circle,draw=black] (pL) at (4, -4.5) {$u_L$};
    
    \node[shape=circle,draw=black] (X11) at (-6, -6) {$X_1^{(1)}$};
    \node[] at (-4, -6) {$\cdots$};
    \node[shape=circle,draw=black] (Xm1) at (-2, -6) {$X_1^{(p_1)}$};

    \node[shape=circle,draw=black] (X1L) at (2, -6) {$X_L^{(1)}$};
    \node[] at (4, -6) {$\cdots$};
    \node[shape=circle,draw=black] (XmL) at (6, -6) {$X_L^{(p_L)}$};

    \path [->](topA) edge node[left] {} (A1);
    \path [->](topA) edge node[left] {} (AL);
    
    \path [->](topp) edge node[left] {} (A1);
    \path [->](topp) edge node[left] {} (AL);
    
    \path [->](A1) edge node[left] {} (X11);
    \path [->](A1) edge node[left] {} (Xm1);
    
    \path [->](p1) edge node[left] {} (X11);
    \path [->](p1) edge node[left] {} (Xm1);
    
    \path [->](AL) edge node[left] {} (X1L);
    \path [->](AL) edge node[left] {} (XmL);
    
    \path [->](pL) edge node[left] {} (X1L);
    \path [->](pL) edge node[left] {} (XmL);
\end{tikzpicture}
\caption{Representation of model~\eqref{eq:hierarchical} as a directed acyclic graph.}
\label{fig:graph}
\end{figure}
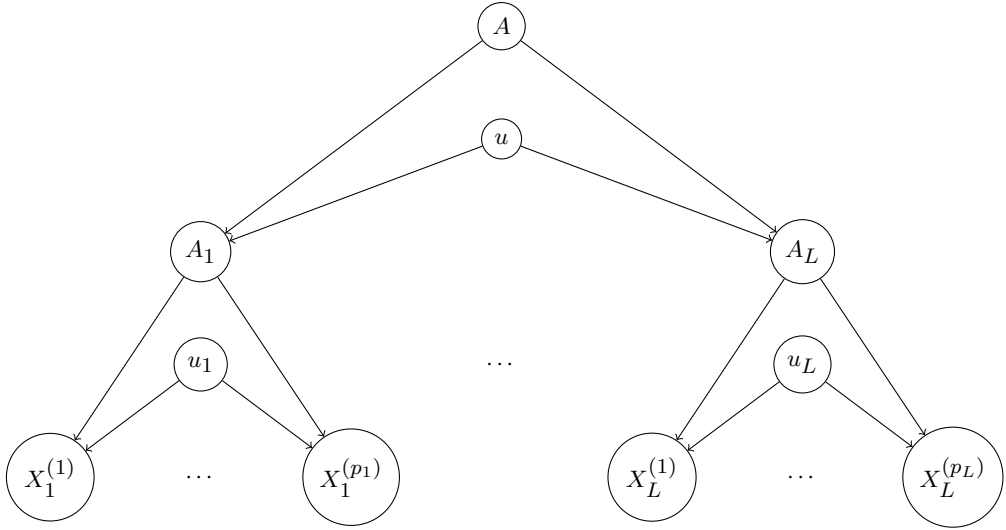

Let~$(P_{A, u})$ be a family of Hamming-based probability distributions.
We propose the following hierarchical model
\begin{equation}\label{eq:hierarchical}
  \left\{
    \begin{array}{lcllll}
      A_i & \lvert & A, \, u & \sim & P_{A, u}, & \quad \mathrm{for} \ 1 \leq i \leq L, \\
      X_i^{(j)} & \lvert & A_i, \, u_i & \sim & P_{A_i, u_i}, & \quad \mathrm{for} \ 1 \leq i \leq L, \quad \mathrm{and} \ 1 \leq j \leq p_i,
    \end{array}
  \right.
\end{equation}
with parameters~$A \in \Pn$,~$u \in I$, and~$u_1, \dots, u_L \in I$. This model is illustrated by Figure~\ref{fig:graph}.
We~will use a product prior for the parameters.

\begin{rem}
To simplify the presentation, we assume that the families of probability distributions for~$A_i$ and~$X_i^{(j)}$ are the same.
What follows can be transparently generalized to the case where these families are different.
\end{rem}

\subsection{Fisher's noncentral hypergeometric distribution}\label{sec:fisher_example}

The results of statistical inference are more easily interpreted with the following family of probability distributions.

Let $I = \left[0, \, + \infty\right)$ and, for $u \in I$, define $e_{u}$ as Fisher's noncentral hypergeometric distribution
\begin{equation}\label{eq:fhg}
e_{u}(k) 
\propto
\binom{N}{k} \binom{n}{n - k} u^k,
\quad
k \in \{0, \dots, n\}.
\end{equation}
Let $0 \leq p_1, p_2 \leq 1$
and
\begin{equation}\label{eq:omega}
u = \frac{p_1 (1 - p_2)}{(1 - p_1) p_2}.
\end{equation}
Then, Algorithm~\ref{alg:fhg} generates samples from~$P_{A, u}$.
\begin{algorithm}
\caption{
Generating samples according to~$P_{A, u}$ when $e_{u}$ is Fisher's noncentral hypergeometric distribution with $u$ given by~\eqref{eq:omega}, for~$0 \leq p_1, p_2 \leq 1$.}\label{alg:fhg}
\begin{algorithmic}
\State {\bfseries Input:} $0 \leq p_1, p_2 \leq 1$
\State $X = \{ \}$
\While{$\# X \neq n$}
\State $X = \{ \}$
\For{$a \in A$}
\State Draw $U \sim \mathrm{Bernoulli}\left( p_1\right)$
\If{$U = 1$}
    \State $X \gets X \cup \{ a \}$
\EndIf 
\EndFor
\For{$a \in A^c$}
\State Draw $U \sim \mathrm{Bernoulli}\left( p_2\right)$
\If{$U = 1$}
    \State $X \gets X \cup \{ a \}$
\EndIf 
\EndFor
\EndWhile
\end{algorithmic}
\end{algorithm}

An interesting feature of Hamming-distance probability distributions
with Fisher's noncentral hypergeometric distribution is that,
for each~$u \geq 0$, there exists a positive constant~$C(u)$ such that, for all~$X \in \Pn$, we have:
\begin{equation}\label{eq:hist_prop_single}
P_{A, u}(X) = C(u) u^k, \quad k = \left( X \cap A^c \right).
\end{equation}
Thus, the probability distribution~$P_{A, u}$ is Hamming decreasing, for~$u \in (0, \, 1)$.
Returning to~\eqref{eq:omega}, it can be seen it amounts to assume that~$p_1 > p_2$:
that is to say that the elements of~$A$ have more probability than that of~$A^c$.
Furthermore,
the probability distribution~$P_{A, 1}$ is uniform on~$\Pn$ and~$P_{A, 0}$ puts all its mass at~$A$.
Observe the analogy with location–scale families of distributions.
(However, this analogy does not hold when~$u > 1$ since~$A$ minimizes~$X \mapsto P_{A, u}(X)$.)

Actually, for~$u \leq 1$, Fisher’s noncentral hypergeometric distribution
even satisfies a stronger property.
Indeed, writing~$A = \{a_1, \dots, a_n\}$ and using~\eqref{eq:hist_prop_single}, we have
\begin{align}\label{eq:prop_fhg}
\mathcal{L} \left( X_1, \dots, X_p; \, A, \, u \right)
%& \, = p \ln \left( C(u) \right) + \ln (u) \sum_{i=1}^p k_i \\
& \, = p \ln \left( C(u) \right) + n p \ln (u) - \ln(u) \sum_{j=1}^n \# \{ i \colon a_j \in X_i \}.
\end{align}
Therefore, the likelihood is purely a function of the sum of the numbers of times the elements of~$A$
appear in the data:
it does not depend on whether or not elements have been selected simultaneously.
Returning to the example from Section~\ref{sec:setup},
we see that, for~$u < 1$, the set \hbox{$A = \{1, 2, 3\}$} chosen corresponds to the maximum likelihood estimate
under Fisher’s noncentral hypergeometric distribution.

\begin{figure}
\centering
\scalebox{0.95}{\input{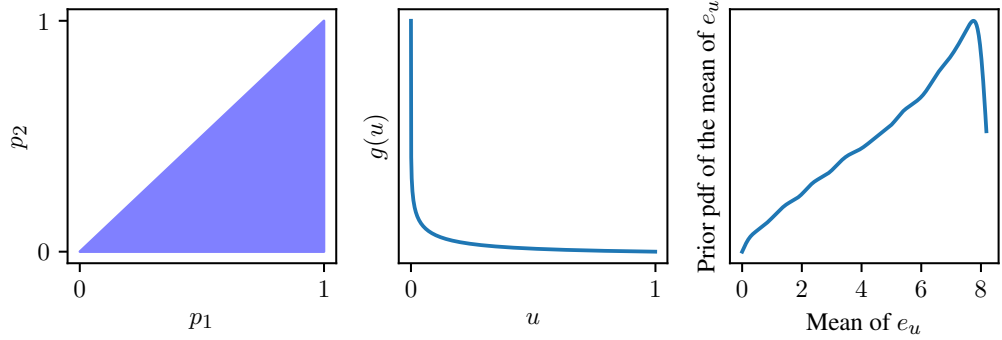}}
\caption{Left: uniform prior on~$\{ 0 < p_2 \leq p_1 < 1\}$.
Middle: the resulting prior probability density on~$u$.
Right: kernel density estimate of the resulting prior probability distribution
function on the mean~$\sum_{k=0}^n k \cdot e_u(k)$ of~$e_u$.}
\label{fig:pdf_omega}
\end{figure}

For the parameter~$u$, we propose a uniform prior on the triangle~$\{ 0 < p_2 \leq p_1 < 1\}$
which we push forward to a prior supported on~$(0, \, 1]$ for~$u$ using~\eqref{eq:omega}.
A straightforward change of variable shows that the resulting prior has density
\begin{equation}\label{eq:prior_omega}
g(u) = 2 \int_0^1 \frac{s - s^2}{(1 - s + u s)^2} \mathrm{d} s
= \frac{
4 (1 - u) + 2 (u + 1) \ln(u)
}{
\left( u - 1 \right)^3
}, \quad u \in (0, \, 1].
\end{equation}
See Figure~\ref{fig:pdf_omega} for an illustration.

% prior sur ω discuter sur son choix et son influence. perspective? on en a esseyé plusieur avec le même support et le résultat est le même. dire que infini en 0 n'est pas un problème. Quid de jeu de données et 

\subsection{Binomial distribution}\label{sec:binomial}

Let $I = [0, \, 1]$ and $e_u = \mathrm{Bin} \left(n, \, u\right)$, for $u \in I$.
In this case, for~$A \in \Pn$, repeating~$n$ times the following procedure generates samples from~$P_{A, u}$.
Draw~$Z$ according to a Bernoulli distribution with parameter~$u$. If~$Z = 1$, then draw without replacement
an element from~$A^c$ uniformly, else, draw without replacement an element from~$A$ uniformly.

Does~$P_{A, u}$ satisfies Definition~\ref{def:gaussian_prop}? If~$u$ is close to one, then not, since
it can be seen that drawing the set~$A$ is rare, in this case. The restriction~$u \leq N / (N + n)$ seems reasonable
since it enforces that the elements of~$A$ are more probable in the first step of the procedure described above.
In this case, it is possible to prove the following proposition.
\begin{prop}
Let~$A \in \Pn$.
If $u < N/(N + n)$, then $A$ is the unique mode of~$P_{A, u}$.
Furthermore, if~$u \leq 1/2$ and~$n \leq N / 2$, then~$P_{A, u}$ is Hamming decreasing.
\end{prop}
It can be seen empirically that~$P_{A, u}$ may not be
Hamming non-increasing, when~$u > 1/2$.

Regarding the prior distribution for the parameter~$u$, a beta
distribution---possibly truncated to enforce, e.g.,~$u \leq N/(N + n)$---can
be used, for instance.

For the rest of this article, we shall mostly focus on Fisher's noncentral hypergeometric distribution,
with~$I = (0, \, 1]$.

\section{Inference}
\label{sec:inference}
\subsection{Inference for a one-stage model}\label{sec:inf-no_lab_effect}

Suppose the data $\underline{X} = \left(X_1, \dots, X_p \right) \in \Pn^p$
are organized without subgroups of operators from different laboratories.
The task is to infer the parameters~$A$ and~$u$ of~$P_{A, u}$.

\begin{figure}
\centering
\scalebox{0.9}{\input{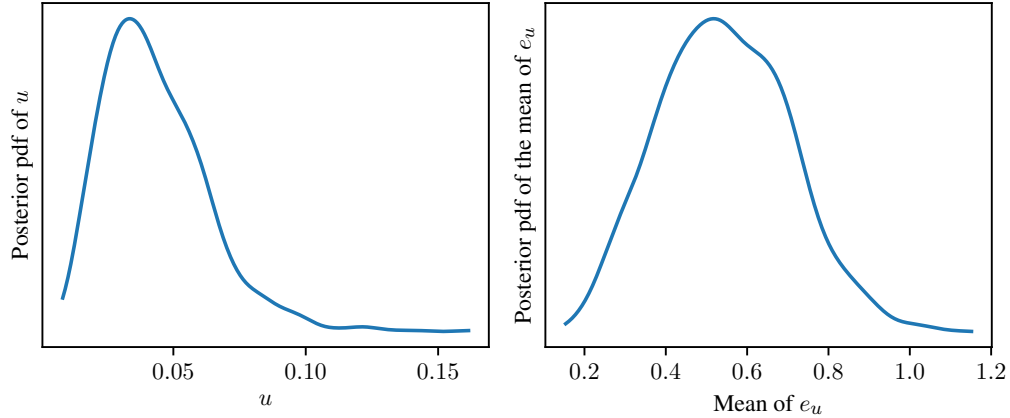}}
\caption{Kernel density estimates of the posterior probability distribution function of~$u$
using $10^3$ posterior samples
obtained using the brute-force scheme on the data from~Table~\ref{tab:example}
with Fisher's noncentral hypergeometric distribution, a uniform prior for~$A$,
and the prior~\eqref{eq:prior_omega} for~$u$.
A uniform grid of size~$10^4$ was used to invert the posterior cumulative distribution function of~$u$.
Left: posterior probability distribution function of~$u$.
Right: posterior probability distribution function of the mean of~$e_u$.}
\label{fig:posterior_toy_example}
\end{figure}

A basic three-steps brute-force scheme will serve as a baseline.
More precisely, we compute the evidence using a Monte Carlo
prior sample of~$u$ and a loop over all~$\Pn$.
Then, we loop again over~$\Pn$ and keep only those~$A \in \Pn$
having their marginal likelihood Monte Carlo estimate~$\hat{P}(\underline{X} \, \lvert \, A)$
such that~$\# \Pn \cdot \hat{P}(\underline{X} \, \lvert \, A)$
is larger than~$\epsilon$ times the evidence, for a prescribed numerical tolerance~$\epsilon > 0$.
Finally, we can repeatedly sample from the posterior distribution by alternating
sampling from the truncation of the posterior of~$A$
and sampling the posterior distribution of~$u$ given~$\underline{X}$ and~$A$.
This latter operation is performed by numerically inverting the cumulative distribution function.
% described by Algorithm~\ref{alg:brute-force_no-op}.
% \begin{algorithm}
% \caption{Basic brute-force inference scheme.
% Two loops are used to save memory.
% The Log-Sum-Exp-Trick may be used to avoid underflow.}\label{alg:brute-force_no-op}
% \begin{algorithmic}
% \State {\bfseries Input:} Data $\underline{X}$;
% a parametrized family~$(P_{A, u})$ in the sense of Definition~\ref{def:loc_scale_dist};
% a prior~$p(u)$;
% and a numerical tolerance~$\epsilon$
% \State Sample~$u_1, \dots, u_q \sim p(u)$
% \State $\hat{P}( \underline{X} ) = 0$
% \For{$A \in \Pn$}
% \State $\hat{P}( \underline{X} ) \gets \hat{P}( \underline{X} ) + q^{-1} \sum_{i=1}^q P(\underline{X} \, \lvert \, A, \, u_i)$
% \EndFor 
% \State  $A_{\mathrm{pos}} = []$
% \State  $P_{\mathrm{pos}} = []$
% \For{$A \in \Pn$}
% \State $\hat{P}(\underline{X} \, \lvert \, A) \gets q^{-1} \sum_{i=1}^q P(\underline{X} \, \lvert \, A, \, u_i)$
% \If{$\# \Pn \cdot \hat{P}(\underline{X} \, \lvert \, A) > \epsilon \hat{P}( \underline{X} )$}
%    \State Append $A$ to~$A_{\mathrm{pos}}$
%  	\State Append $\hat{P}(\underline{X} \, \lvert \, A)$ to~$P_{\mathrm{pos}}$
% \EndIf
% \EndFor 
% \end{algorithmic}
% \end{algorithm}
Using this brute-force scheme on the data from~Table~\ref{tab:example}
with Fisher's noncentral hypergeometric distribution, a uniform prior for~$A$, and the prior~\eqref{eq:prior_omega} for~$u$
yield a posterior probability of~$1 - 10^{-8}$ for~$A = \{ 1, 2, 3 \}$ and posterior samples of~$u$
shown in Figure~\ref{fig:posterior_toy_example}.
The posterior samples can be used to detect atypical responses among the~$X_i$s
by computing, for each~$i$, the empirical posterior average of the
probability
\begin{equation}\label{eq:prob_larger_deviation}
\sum_{k = \#(X_i \cap A^c)}^n e_{u}(k)
\end{equation}
of having more elements outside~$A$ than~$X_i$. 
These numbers can be interpreted as posterior $p$-values.
%\begin{table}\label{tab:example_statistical_analysis}
%\caption{Estimates of the posterior p-values~\eqref{eq:prob_larger_deviation}, for the data shown in Table~\ref{tab:example}.}
%\centering
%\begin{tabular}{| c || c | c | c | c | c | c | c | c | c | c | c | c |}
%\hline
%$X_i$ & $X_{1}$ & $X_{2}$ & $X_{3}$ & $X_{4}$ & $X_{5}$ & $X_{6}$ & $X_{7}$ & $X_{8}$ & $X_{9}$ & $X_{10}$ & $X_{11}$ & $X_{12}$\\
%\hline
%Estimate & $1.0$ & $1.0$ & $1.0$ & $\approx 0.48$ & $1.0$ & $1.0$ & $\approx 0.48$ & $1.0$ & $\approx 0.48$ & $1.0$ & $1.0$ & $1.94 \cdot 10^{-3}$\\
%\hline
%\end{tabular}
%\end{table}
Computing them
for the data from~Table~\ref{tab:example} yields an empirical posterior average
of~$0.2 \%$ for~$X_{12}$, which is thus found to have significantly more
deviations from the central set under the model.

The approach presented above is not scalable since~$\# \Pn = \binom{M}{n}$, with~$M = \# \Ocal$.
We also propose an MCMC approach using a Metropolis-within-Gibbs sampler.
Specifically,
a new element of~$I$ is proposed using a Gaussian random walk on some
reparametrization (e.g., using a logit).
Then, given~$A \in \Pn $ and~$u \in I$, a new central set is proposed by
drawing uniformly an~$a \in A$ and replacing it with an element sampled
from~$\{ a \} \cup (\Ocal \setminus A)$---with probabilities to be specified.

\subsection{Inference for a two-stage model}\label{sec:inf-lab_effect}

Consider the case where the data are presented as in Section~\ref{sec:lab_effect}.
An inference method has been developed specifically for the EMC interlaboratory comparison.
This method shares similarities with the brute-force scheme presented in Section~\ref{sec:inf-no_lab_effect},
but involves some calculation tricks
to take account of the fact that---given the parameters---the choices made by operators
in the same laboratory are no longer assumed to be independent.

Assuming a product prior on~$(A, \, u, \, u_1, \dots, u_L)$,
the method consists in using the ancestral sampling scheme
\begin{equation}\label{eq:ancestral}
  \left\{
    \begin{array}{lclcl}
	  A & \lvert & \ \underline{\mathbb{X}} & & \\
      u & \lvert & \ \underline{\mathbb{X}}, \, A & & \\
      A_i  & \lvert & \ \underline{\mathbb{X}}, \, A, \, u, & \quad \mathrm{for} \ 1 \leq i \leq L, \\
	  u_i  & \lvert & \ \underline{\mathbb{X}}, \, A, \, u, \, A_i, & \quad \mathrm{for} \ 1 \leq i \leq L, \\
    \end{array}
  \right.
\end{equation}
given by the conditional independencies.

\begin{figure}
\centering
\begin{tikzpicture}
    \node[shape=circle,draw=black] (topA) at (0, 0){$A$};
    
    \node[shape=circle,draw=black] (topp) at (0, -1.5){$u$};
    
    \node[shape=circle,draw=black] (A1) at (-4, -3) {$X_1^{(1)}, \dots, X_1^{(p_1)}$};
    \node[] at (0, -3) {$\cdots$};
    \node[shape=circle,draw=black] (AL) at (4, -3) {$X_L^{(1)}, \dots, X_L^{(p_L)}$};
    
    \path [->](topA) edge node[left] {} (A1);
    \path [->](topA) edge node[left] {} (AL);
    
    \path [->](topp) edge node[left] {} (A1);
    \path [->](topp) edge node[left] {} (AL);
    
\end{tikzpicture}
\caption{Representation of model~\eqref{eq:hierarchical} as a directed acyclic graph after marginalizing out the~$A_i$s and the~$u_i$s.}
\label{fig:collapsed_graph}
\end{figure}
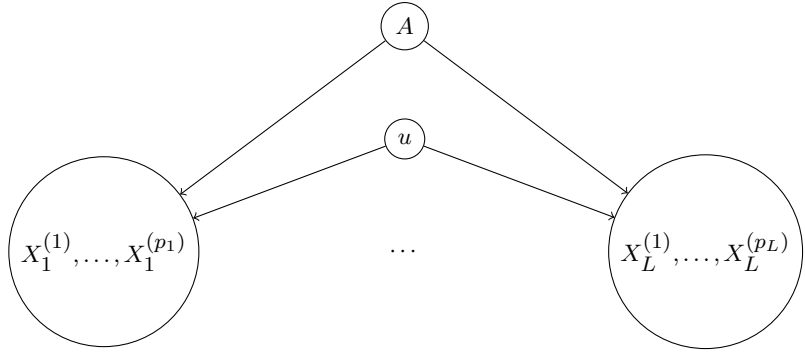

A key step is to marginalize out the~$A_i$s and the~$u_i$s
given~$A$ and~$u$, making it possible to modify the graph shown in Figure~\ref{fig:graph}
to the one shown in Figure~\ref{fig:collapsed_graph}.

For~$p \geq 1$, $(r_1, \dots, r_p) \in \{0, \dots, n\}^p$, and~$Y_1, \dots, Y_p \in \Pn$, let
\begin{equation}\label{eq:def_c_numbers}
\mathcal{C}_n \left( Y_1, \dots, Y_p; \, r_1, \dots, r_p \right)
=
\# \left\{ Y \in \Pn \, \colon \, \# \left( Y \cap Y_i \right) = r_i, \, \forall 1 \leq i \leq d \right\}.
\end{equation}
This value does not depend on~$Y_1$ for~$p = 1$, so we will write~$\mathcal{C}_n \left( r \right) = \mathcal{C}_n \left( Y_1; \, r \right) = \binom{n}{r} \binom{N}{n - r}$.
Writing~$g$ for the prior on~$u_i$, we then have
\begin{align*}
& P \left( X_i^{(1)}, \dots, X_i^{(p_i)} \, \lvert \, A, \, u \right) \\
& \, = \sum_{A_i \in \Pn}
\frac{
e_u \left( \# \left( A_i \cap A^c \right) \right)
\int \prod_{j = 1}^{p_i} e_{u_i} \left( \# \left( X_i^{(j)} \cap A_i^c \right) \right) g(u_i) \mathrm{d}u_i
}{
\mathcal{C}_n \left( n - \# \left( A_i \cap A^c \right) \right)
\prod_{j = 1}^{p_i}
\mathcal{C}_n \left( n - \# \left( X_i^{(j)} \cap A_i^c \right) \right)
}\\
& = \,
\sum_{q = 0}^n \
\sum_{\underline{r} = (r_1, \dots, r_{p_i}) \in \{ 0, \dots, n \}^{p_i}}
\mathcal{C}_n \left( A, \, \underline{X}_i; \, q, \, \underline{r} \right)
\frac{
e_u \left( n - q \right)
\int \prod_{j = 1}^{p_i} e_{u_i} \left( n - r_j \right) g(u_i) \mathrm{d}u_i
}{
\mathcal{C}_n \left( q \right)
\prod_{j = 1}^{p_i}
\mathcal{C}_n \left( r_j \right)
}\\
& = \,
\sum_{q = 0}^n
\frac{
e_u \left( n - q \right)
}{
\mathcal{C}_n \left( q \right)
}
\underbrace{
\sum_{\underline{r} = (r_1, \dots, r_{p_i}) \in \{ 0, \dots, n \}^{p_i}}
\mathcal{C}_n \left( A, \, \underline{X}_i; \, q, \, \underline{r} \right)
\frac{ \int \prod_{j=1}^{p_i} e_{u_i} \left( n - r_j \right) g(u_i) \mathrm{d}u_i}{
\prod_{j=1}^{p_i} \mathcal{C}_n \left( r_j \right)
}}_{D(q, \, i, \, A)}.
\end{align*}

The brute-force scheme then consists in sampling~$u^{(1)}, \dots, u^{(N)}$ according to the prior on~$u$
and looping over $A \in \Pn$ to compute the empirical estimates
\begin{equation}\label{eq:approx_margin_lik_A_op}
\frac{1}{N} \sum_{j=1}^N  \prod_{i = 1}^L P \left( X_i^{(1)}, \dots, X_i^{(p_i)} \, \lvert \, A, \, u^{(j)} \right).
\end{equation}
This involves pre-calculating the quantities~$D(q, \, i, \, A)$,
%\begin{equation}\label{values_to_be_precomputed}
%\end{equation}
for all~$0 \leq q \leq n$, all~$A \in \Pn$,
and all~$1 \leq i \leq L$.
For this purpose, the Supplementary Material proposes an algorithm based on
hash tables \cite[see, e.g.,][Section~III.11]{cormen2022introduction}.
This algorithm does not scale,
with in particular the number of operators being a strong limiting factor.

Proceeding as described in Section~\ref{sec:inf-no_lab_effect},
one can then use~\eqref{eq:approx_margin_lik_A_op} in a double loop to obtain
an approximation of the posterior distribution of~$A$.
Given a posterior sample of~$A$, one can invert numerically the cumulative distribution function
to sample~$u$. The Supplementary Material
concludes by explaining how to sample~$A_i$ given~$A$ and~$u$.
Finally, given~$A_i$, one can sample~$u_i$ using again the cumulative distribution function.

\section{Application}
\label{sec:application}

\subsection{Case study}\label{sec:case_description}

Our work was initiated in the context of an interlaboratory comparison campaign, organized by the
Electromagnetic Compatibility (EMC) Working Group of \emph{Eurolab France}, an association of certified laboratories.  

\begin{figure}
\centering
\includegraphics[scale=0.45]{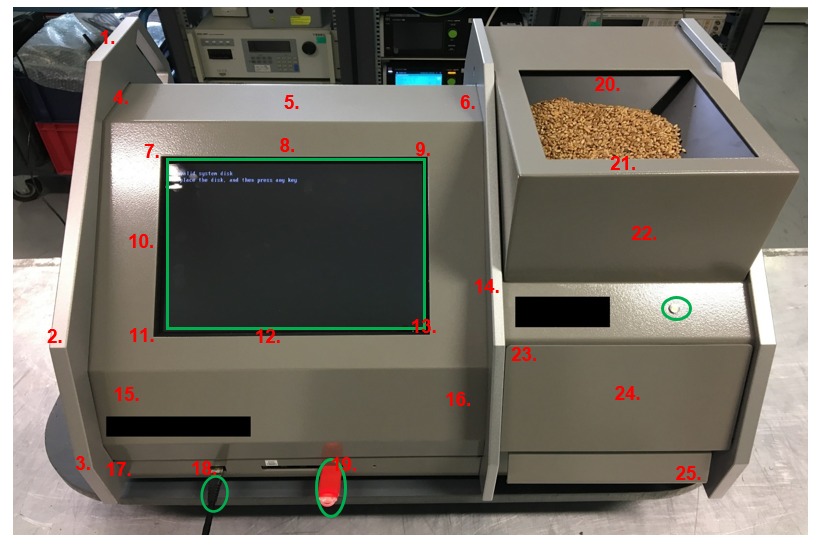}
\includegraphics[scale=0.45]{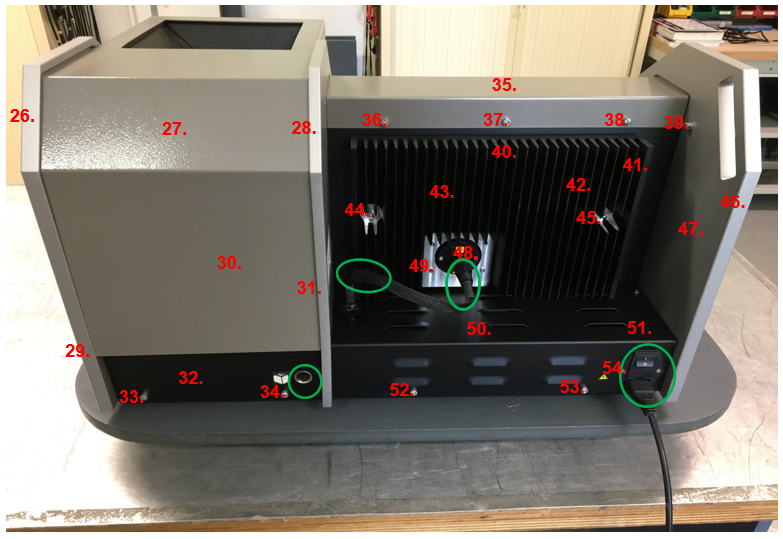}
\caption{Photos given to the participants of the interlaboratory comparison campaign. The injection points to select are indicated on the tested device (an infrared grain analyzer).}
\label{fig:machine}
\end{figure}

The aim of this study is to compare the results of different laboratories when selecting~$10$
electrostatic discharge injection points on an electrical device according to the norm EN~61000-4-2\string:2009.
Each laboratory received a spreadsheet containing information, photos, and appendices to guide
them in selecting injection points. The photos of the device are given in Figure~\ref{fig:machine}.

We apply our proposed method to determine which selection is consensual, and which results deviate
statistically from this consensus. These analyses also enable us to determine whether inter-operator variability is significant,
and whether it contributes to the differences observed between laboratories.

\begin{figure}
\centering
\scalebox{0.65}{\input{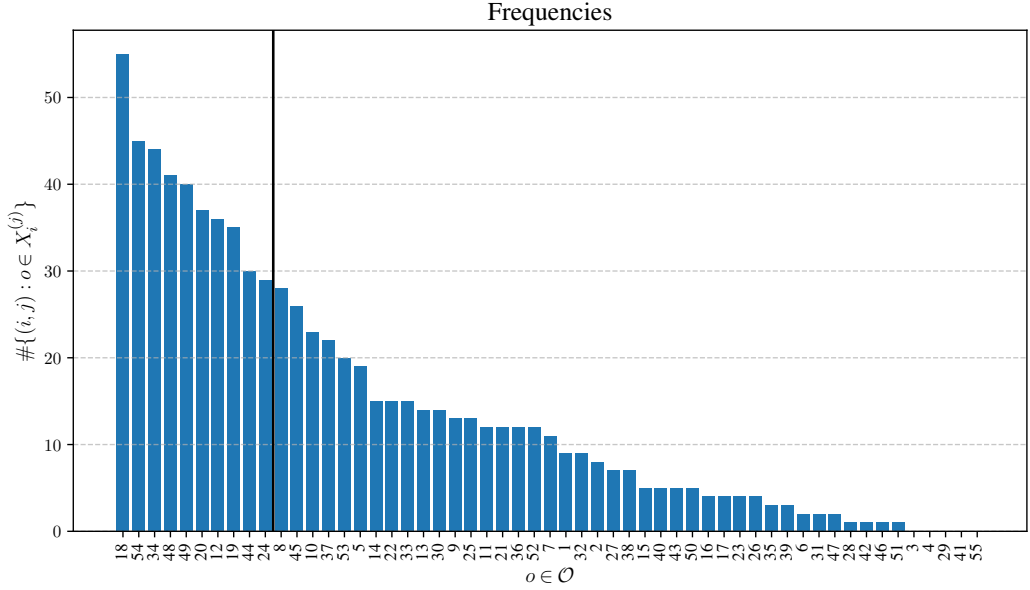}}
\caption{The number of~$X_i^{(j)}$s containing each element of~$\Ocal$.
The black vertical line marks the~$n = 10$ most frequently selected points.}
\label{fig:hist_cil_cem}
\end{figure}

Regarding the data, the convention is that~$\Omega = \{1, \dots, 55\}$ and the choices
$$
\underline{\mathbb{X}} = \left(\underline{X}_1, \dots, \underline{X}_L \right), \quad
\mathrm{with} \ \underline{X}_i = (X_i^{(1)}, \dots, X_i^{(p_i)}) \in \Pn^{p_i}, \quad
\mathrm{for} \ 1 \leq i \leq L,
$$
from~$\sum_{i=1}^L p_i = 78$ operators from~$L = 26$ laboratories were collected.
Figure~\ref{fig:mozaique_cil_cem} summarizes the data.
As for the introductory example, Figure~\ref{fig:hist_cil_cem} represents the number of~$X_i^{(j)}$s containing each element of~$\Ocal$.

\subsection{Statistical analysis, all operators pooled together}\label{sec:app-no_op}

The data is first analyzed statistically by pooling all operators together
in a one-stage model with
Fisher’s noncentral hypergeometric distribution for~$e$,
a uniform prior for~$A$,
and the prior~\eqref{eq:prior_omega} for~$u$.
In this context, we write~$\underline{X} = (X_1, \dots, X_p)$, with~$p = 78$, for the data.
The numbering of the operators is the one appearing on the right-side of Figure~\ref{fig:mozaique_cil_cem}.

\begin{figure}
\centering
\scalebox{0.9}{\input{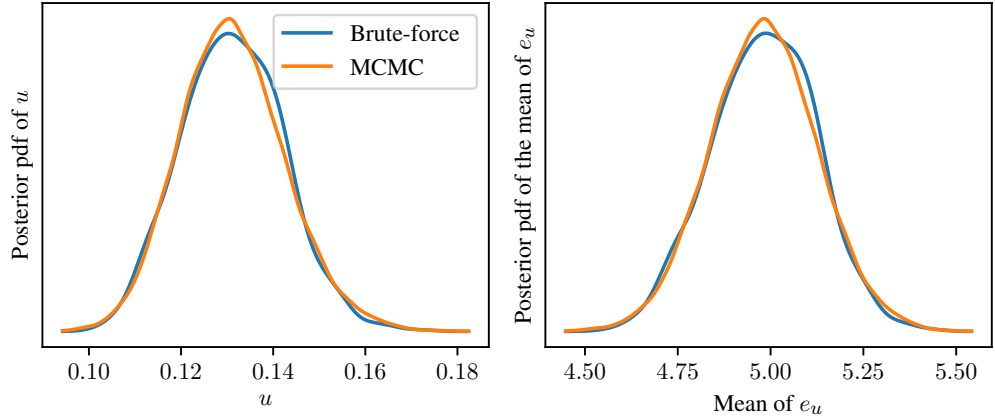}}
\caption{
Kernel density estimates of the posterior probability
distribution functions
of~$u$
obtained using samples generated by the brute-force scheme and the MCMC algorithm
with a one-stage model on the data shown in Figure~\ref{fig:mozaique_cil_cem}.
A uniform grid of size~$10^4$ was used to numerically invert the cumulative distribution function to generate~$10^3$ samples
with the brute-force scheme.
For the MCMC, all the repetitions led to similar kernel density estimates so the results are shown for only one repetition.
A subsampling of~$1$ out of~$46$ samples is performed to reduce autocorrelation in the chain.
Left: posterior probability distribution function of~$u$.
Right: posterior probability distribution function of the mean of~$e_u$.
}
\label{fig:posterior_cil_cem}
\end{figure}

The approximate marginal posterior distributions of~$A$ and~$u$ given by the brute-force
scheme and the MCMC scheme presented in Section~\ref{sec:inf-no_lab_effect} are shown in
Table~\ref{tab:cil_cem_brute_force_statistical_analysis}
and Figure~\ref{fig:posterior_cil_cem}.
Observe the good agreement between the results from the two techniques.
For the brute-force scheme, the value~$\epsilon = 1\%$ was used
and the estimates of the marginal likelihood of~$A$
were computed using~$1000$ samples of~$u$.
For the MCMC algorithm, tuning the variance~$\sigma^2$
of a Gaussian random walk proposal led to a value of~$\sigma^2 = 1/2$,
yielding a~$17 \%$ acceptance rate for the moves along~$u$.
For the proposal distribution for~$A$, given~$a \in A$,
the probabilities of replacing~$a$ by~$o \in \{ a \} \cup (\Ocal \setminus A)$
are chosen proportional to~$\# \{ i \, \colon o \in X_i \} + 1$.
This choice was found to improve the mixing of the chain.
In this setup, we run~$30$ repetitions of~$10^6$ runs of the MCMC algorithm.
A burn-in phrase of~$10^5$ first samples was discarded out.

The approximate posterior distribution of~$A$ given by
Table~\ref{tab:cil_cem_brute_force_statistical_analysis}
is consistent with~\eqref{eq:prop_fhg}.
Indeed, the posterior mode is the set of the $10$ most selected items,
as shown in Figure~\ref{fig:hist_cil_cem}.
The second and third most probable sets correspond, 
respectively, to the replacement of the tenth and ninth most 
chosen elements with the eleventh,
which is chosen only slightly less often.

\begin{table}
\caption{Approximations of the posterior distribution~$P ( A \, \lvert \, \underline{X})$ of~$A$
with a one-stage model for the data shown in Figure~\ref{fig:mozaique_cil_cem}.
For clarity, three sets representing~$99 \%$ of the posterior mass are shown.
For the MCMC, average values over the~$30$ repetitions are presented and standard deviations are shown in parentheses.
Colors are used to highlight elements that do not appear in all the three most probable sets.}
\label{tab:cil_cem_brute_force_statistical_analysis}
\centering
\begin{tabular}{| c | c | c |}
\hline
$A$ & Brute-force & MCMC \\
\hline
$\{12, 18, 19, 20, {\color{blue} 24}, 34, {\color{red} 44}, 48, 49, 54\}$ & $87.228 \%$ & $ 86.920 \% \ (0.690 \%)$ \\
\hline
$\{{\color{orange} 8}, 12, 18, 19, 20, 34, {\color{red} 44}, 48, 49, 54\}$ & $11.297 \%$ & $ 11.319 \% \ (0.640 \%)$ \\
\hline
$\{{\color{orange} 8}, 12, 18, 19, 20, {\color{blue} 24}, 34, 48, 49, 54\}$ & $1.475 \%$ & $ 1.520 \% \ (0.173 \%)$ \\
\hline
\end{tabular}
\end{table}

The NF-ISO-13528 standard for continuous data stipulates that an
``alert'' signal is issued for an absolute z-score between two and three.
An ``action'' signal is issued when this absolute value is greater than three.
Corresponding signals can be issued for set-valued data by using the 
empirical posterior averages of the
$p$-values~\eqref{eq:prob_larger_deviation},
which are shown in Table~\ref{tab:cil_cem_emitted_signals} for both the brute-force scheme
and the MCMC algorithm.
The results are again consistent for the two inference methods.
Observe that a clear decision rule emerges,
with eight elements outside the posterior mode~$\hat{A}$ of~$A$ leading to an alert signal,
and nine or ten elements outside it leading to an action signal.

\begin{table}
\caption{Empirical posterior averages of~\eqref{eq:prob_larger_deviation} for the data shown in Figure~\ref{fig:mozaique_cil_cem}.
The choices are referenced by the numbers shown on the right-hand side of Figure~\ref{fig:mozaique_cil_cem}.
The number of elements outside the posterior mode~$\hat{A}$ are given in the second column.
For the MCMC, average values over the~$30$ repetitions are presented and standard deviations are shown in parentheses.}
\label{tab:cil_cem_emitted_signals}
\centering
\begin{tabular}{| c || c | c | c |}
\hline
$X_i$ & $\# (X_i \cap \hat{A}^c)$ & Brute-force & MCMC \\
\hline
\hline
$X_{1}$ & $8$ & $2.642 \%$ & $ 2.633 \% \ (0.018 \%)$ \\
\hline
$X_{2}$ & $8$ & $2.211 \%$ & $ 2.231 \% \ (0.013 \%)$ \\
\hline
$X_{31}$ & $8$ & $2.642 \%$ & $ 2.633 \% \ (0.018 \%)$ \\
\hline
$X_{43}$ & $8$ & $2.211 \%$ & $ 2.231 \% \ (0.013 \%)$ \\
\hline
$X_{58}$ & $8$ & $2.488 \%$ & $ 2.491 \% \ (0.002 \%)$ \\
\hline
$X_{74}$ & $8$ & $2.211 \%$ & $ 2.231 \% \ (0.013 \%)$ \\
\hline
$X_{75}$ & $8$ & $2.211 \%$ & $ 2.231 \% \ (0.013 \%)$ \\
\hline
\hline
$X_{17}$ & $9$ & $0.283 \%$ & $ 0.284 \% \ (0.001 \%)$ \\
\hline
$X_{32}$ & $10$ & $0.051 \%$ & $ 0.049 \% \ (0.002 \%)$ \\
\hline
$X_{33}$ & $9$ & $0.319 \%$ & $ 0.319 \% \ (0.004 \%)$ \\
\hline
$X_{72}$ & $9$ & $0.249 \%$ & $ 0.252 \% \ (0.002 \%)$ \\
\hline
$X_{73}$ & $9$ & $0.283 \%$ & $ 0.284 \% \ (0.000 \%)$ \\
\hline
\hline
Others & $\leq 7$ & $\geq 5 \%$ & $\geq 5 \%$ \\
\hline
\end{tabular}
\end{table}

\subsection{Statistical analysis taking into account a possible within-laboratory effect}\label{sec:app-op}

The hierarchical model introduced in Section~\ref{sec:lab_effect} is
used to take a possible within-laboratory effect into account.
We will use the notation from Section~\ref{sec:case_description} for the data.
Again, Fisher’s noncentral hypergeometric distribution is used for both stages of the model.
We use a product prior on~$(A, \, u, \, u_1, \dots, u_L)$,
with a flat prior on~$A$ and the same prior~\eqref{eq:prior_omega} for the dispersion parameters.

\begin{figure}
\centering
\scalebox{0.95}{\input{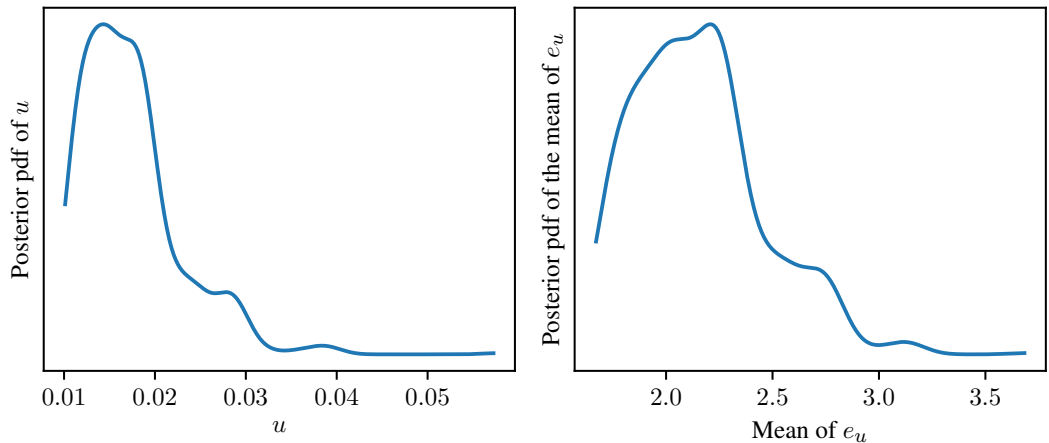}}
\caption{
Kernel density estimate of the posterior probability
distribution function
of~$u$
obtained using a two-stage model and the brute-force scheme presented
in Section~\ref{sec:inf-lab_effect} on the data shown in Figure~\ref{fig:mozaique_cil_cem}.
Left: posterior probability distribution function of~$u$.
Right: posterior probability distribution function of the mean of~$e_u$.
}
\label{fig:posterior_omega_cil_cem_op}
\end{figure}

The inference is made by running the algorithm described in Section~\ref{sec:inf-lab_effect}
and in the Supplementary Material. The resulting marginal posteriors of~$A$ and~$u$
are shown in Table~\ref{tab:cil_cem_brute_force_statistical_analysis_op} and in Figure~\ref{fig:posterior_omega_cil_cem_op}.

\begin{table}
\caption{Approximations of the posterior distribution~$P ( A \, \lvert \, \underline{\mathbb{X}})$
of~$A$, taking into account a possible within-laboratory effect.
For clarity, four sets representing~$99 \%$ of the posterior mass are shown.
Colors are used to highlight elements that do not appear in all the four most probable sets.}
\centering
\begin{tabular}{| c | c |}
\hline
$A$ & Brute-force \\
\hline
$\{{\color{orange} 8}, 12, 18, 19, 20, 34, {\color{red} 44}, 48, 49, 54\}$ & $58.099 \%$ \\
\hline
$\{12, 18, 19, 20, 34, {\color{red} 44}, {\color{olive} 45}, 48, 49, 54\}$ & $36.491 \%$ \\
\hline
$\{12, 18, 19, 20, {\color{blue} 24}, 34, {\color{red} 44}, 48, 49, 54\}$ & $4.587 \%$ \\
\hline
$\{{\color{orange} 8}, 12, 18, 19, 20, {\color{blue} 24}, 34, 48, 49, 54\}$ & $0.823 \%$ \\
\hline
\end{tabular}
\label{tab:cil_cem_brute_force_statistical_analysis_op}
\end{table}

Observe the difference between the posterior distributions of~$A$ shown in Table~\ref{tab:cil_cem_brute_force_statistical_analysis}
and Table~\ref{tab:cil_cem_brute_force_statistical_analysis_op}. The posterior mode with the one-stage model---which
is also the~$10$ elements most often chosen by operators---has~$4.587 \%$ posterior probability
under the two-stage model. The other set from Table~\ref{tab:cil_cem_brute_force_statistical_analysis}
containing~$24$ also has lower posterior probability.
Actually, a set containing~$45$---and not appearing in Table~\ref{tab:cil_cem_brute_force_statistical_analysis}---has now
substantially more posterior probability than the two containing~$24$. Observe in Figure~\ref{fig:hist_cil_cem}
that~$45$ is less chosen. However, inspection of Figure~\ref{fig:mozaique_cil_cem} reveals that~$24$
is chosen by the majority of operators from laboratory~$10$ and laboratory~$25$,
which we shall see have substantial posterior probabilities of having intra-laboratory consensuses statistically distant from
the rest of the participants. In contrast, the items~$8$ and~$45$ are chosen less often, overall, but more by laboratories
making more consensual choices.

The interval~$[0.01, \, 0.03]$ contains~$99 \%$ of the posterior mass of~$u$, showing that
the model is confident about the existence of laboratory-specific biases.
To test for the existence of a within-laboratory effect, we
compute the Bayes factor which is the ratio between the evidence~$P \left( \underline{\mathbb{X}} \right)$
of the model taking a possible within-laboratory effect into account and
the evidence~$P \left( \underline{X} \right)$ of the model from Section~\ref{sec:app-no_op},
which pools all the operators together.
The evidences are estimated when applying the brute-force inference methods, giving
$$
\frac{P \left( \underline{\mathbb{X}} \right)}{P \left( \underline{X} \right)}
\approx \frac{ \mathrm{exp} \left\lbrace -1554.4 \right\rbrace}{ \mathrm{exp} \left\lbrace -1616.0 \right\rbrace}
\approx 6 \cdot 10^{26}.
$$
According to the classification proposed by~\citet[p.~9]{kass1995bayes},
this gives a very substantial indication that the
laboratory-effect model is better supported by the data.

Consider the problem of detecting outlier laboratories, i.e.,
laboratories whose intra-laboratory consensus~$A_i$
seems to deviate significantly more from~$A$ than the others.
One can adapt~\eqref{eq:prob_larger_deviation} to consider the posterior distribution of
\begin{equation}\label{eq:prob_larger_deviation_op}
\Gamma_i = \sum_{k = \#(A_i \cap A^c)}^n e_{u}(k)
\end{equation}
given~$\underline{\mathbb{X}}$.
Note that neither~$A$, nor~$u$, nor~$A_i$ are directly observed.

\begin{figure}
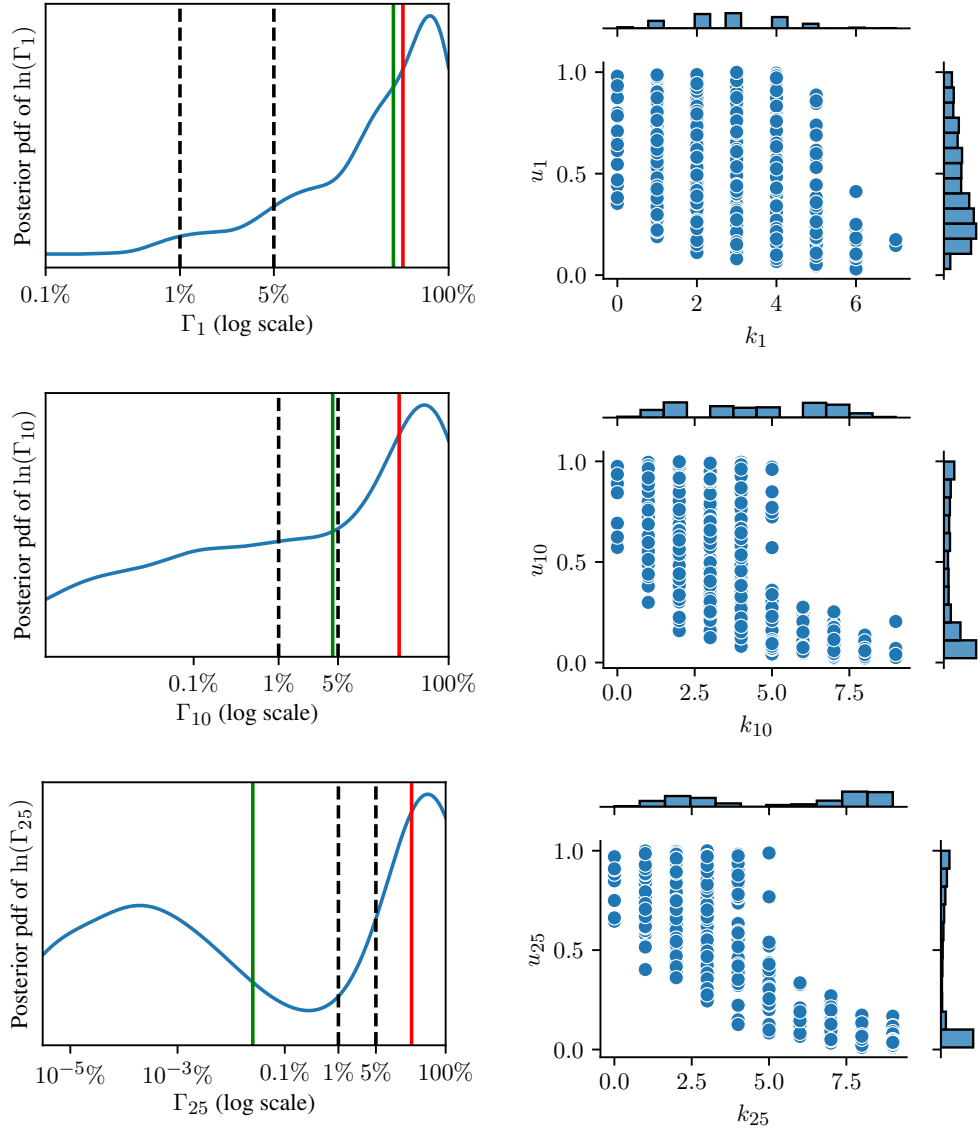

\centering
\scalebox{0.9}{\input{figures/posterior_op_plots/p_val_plot-1.pgf}}
\scalebox{0.9}{\input{figures/posterior_op_plots/k_i_u_i-1.pgf}}
\scalebox{0.9}{\input{figures/posterior_op_plots/p_val_plot-10.pgf}}
\scalebox{0.9}{\input{figures/posterior_op_plots/k_i_u_i-10.pgf}}
\scalebox{0.9}{\input{figures/posterior_op_plots/p_val_plot-25.pgf}}
\scalebox{0.9}{\input{figures/posterior_op_plots/k_i_u_i-25.pgf}}
\caption{
Posterior distributions of~$\Gamma_i$,~$\overline{e}_{i}$,
and~$k_i = \# (A_i \cap A^c)$ for three laboratories.
The left plots show samples of the posterior distribution of~$\ln( \Gamma_i)$ and the right plots show
joint plots of posterior samples of~$(\overline{e}_{i}, \, k_i)$. On the left plots, the red line stands for the logarithm of the posterior mean of~$\Gamma_i$
and the green line stands for the logarithm of the posterior median of~$\Gamma_i$.
The upper plots stand for laboratory~$1$, the middle plots stand for laboratory~$10$,
and the lower plots stand for laboratory~$25$.
}
\label{fig:posterior_cil_cem_op_omega_i_gamma_i_k_i}
\end{figure}

The posterior distributions of the~$\Gamma_i$s are more spread out
than those of~\eqref{eq:prob_larger_deviation}.
Consequently, the decisions depend on the choice of
a point estimate such as the posterior mean or the posterior median.
Laboratory~$1$, laboratory~$10$, and laboratory~$25$
are the three borderline cases that emerge from the analysis.
Observe in Figure~\ref{fig:mozaique_cil_cem} that operators in
these laboratories select more often items outside
the support of the posterior of~$A$.
In the following, write~$\overline{e}_i$ for the mean of~$e_{u_i}$ to simplify notation.
Figure~\ref{fig:posterior_cil_cem_op_omega_i_gamma_i_k_i}
represents the posterior distributions of~$\Gamma_i$,~$\overline{e}_i$,
and~$k_i = \# (A_i \cap A^c)$ for these laboratories.

Within-laboratory dispersion is low in laboratories~$10$ and~$25$:
for each of these two laboratories, operators tend
to make similar non-consensual choices.
Consequently,~$\overline{e}_{10}$
and~$\overline{e}_{25}$ have significant posterior probability masses 
on low numbers of deviations
and, accordingly,~$\Gamma_{10}$ and~$\Gamma_{25}$ have
important probability masses below~$5 \%$.
However,~$\overline{e}_{10}$
and~$\overline{e}_{25}$ also have probability
masses on large values, translating into probabilities of~$k_{10}$ and~$k_{25}$ being low
due to posterior correlation.
This means that there is still a chance that operators have accidentally
made similar errors  in the absence of  bias.
Consequently,~$\Gamma_{10}$ and~$\Gamma_{25}$ have probability masses way above~$5 \%$,
which inflate the posterior means.
However, the posterior median of~$\Gamma_{10}$
is just below~$5 \%$ and
the posterior median of~$\Gamma_{25}$ is lower than~$0.1 \%$.
%Therefore, an alert signal is issued for laboratory~$10$ and an action signal
%is issued for laboratory~$25$.

Regarding laboratory~$1$, the posterior distribution assigns
more mass on large mean numbers~$\overline{e}_1$ of deviations from~$A_1$.
This is expected as Figure~\ref{fig:mozaique_cil_cem} 
shows more diverse selections within this laboratory.
This dispersion makes the results insufficient to conclude about
the existence of a bias in this laboratory since there is only a small probability that $\Gamma_1$ is
below~$5~\%$.
However, the operators from laboratory~$1$ exhibit
higher variability in their choices than is observed overall in the comparison.
Indeed, compare the posterior distribution of~$\overline{e}_{10}$
with the estimated variability in the choices of the operators pooled together,
as shown in Figure~\ref{fig:posterior_cil_cem}.
In particular, the posterior median of~$\overline{e}_{1}$ is about~$6.8$
deviations between an operator from laboratory~$1$ and~$A_1$.
Therefore, the posterior distribution quantitatively confirms a noticeable disparity 
in laboratory~$1$, which must be reported.

\section{Conclusions and perspectives}
\label{sec:conclusions}

% Se défendre sur l'alerte en cas de détection bimodalité 
% Ce modèle gère beaucoup de données possible (coup de choisir 10 parmi 20 points égaux en fréquence)

This article proposes Hamming-based probability distributions over subsets to analyze set-valued data.
A Bayesian approach is used to analyze the results of an
interlaboratory comparison made by the Electromagnetic Compatibility Working Group of Eurolab France.
The existence of a within-laboratory effect is also highlighted using a Bayes factor against a hierarchical model.

Two inference techniques are proposed for one-stage models: a brute-force approach and an MCMC proposal.
For hierarchical models, we propose an adaptation of the
brute-force approach that rely on calculation tricks. However,
it does not scale, in particular with respect
to the number of operators in the laboratories. 
Unfortunately, despite our best efforts (leveraging in particular the state-of-the-art approaches of \citealp{rhodes2022enhanced},
and \citealp{zhang2022langevin}), we did not succeed in making the inference using MCMC for the hierarchical model.
Future work should address this limitation.

Another line of research for future work would be to investigate the choice of the parametrized family~$e\colon I \to E$.
Section~\ref{sec:fisher_example} proposes Fisher's noncentral hypergeometric distribution
to enhance interpretability. However some data sets could be better modeled using other distributions.
In this case, the likelihood would not depend only on the sum of the numbers of times the objects
appear in the data anymore:
it~may also depend on whether or not elements have been selected simultaneously.
Mixtures of families can be used to give more flexibility, but guidance must be
provided to select these families. It is plausible that restricting the choice to families
satisfying Definition~\ref{def:gaussian_prop} could be reasonable for most applications.

Finally, some work is probably required on decision rules to detect outliers.
For instance, in Section~\ref{sec:app-op},
due to the spread of the posterior distribution for the hierarchical model,
a visual exploration is made to detect outliers.
Proposing Bayesian tests with improved error rates can
help to better judge the responses of the participants, and therefore, to improve their practices.

%%%%%%%%%%%%%%%%%%%%%%%%%%%%%%%%%%%%%%%%%%%%%%
%% Support information, if any,             %%
%% should be provided in the                %%
%% Acknowledgements section.                %%
%%%%%%%%%%%%%%%%%%%%%%%%%%%%%%%%%%%%%%%%%%%%%%
\begin{acks}[Acknowledgments]
The authors would like to thank the
Electromagnetic Compatibility Working Group of Eurolab France
for permission to show results based on the data from the comparison.
\end{acks}

%%%%%%%%%%%%%%%%%%%%%%%%%%%%%%%%%%%%%%%%%%%%%%
%% Supplementary Material, including data   %%
%% sets and code, should be provided in     %%
%% {supplement} environment with title      %%
%% and short description. It cannot be      %%
%% available exclusively as external link.  %%
%% All Supplementary Material must be       %%
%% available to the reader on Project       %%
%% Euclid with the published article.       %%
%%%%%%%%%%%%%%%%%%%%%%%%%%%%%%%%%%%%%%%%%%%%%%
\begin{supplement}%
\stitle{Inference algorithm for the two-stage model.}%
\sdescription{Details on how to precompute the values of~$D(q, \, i, \, A)$
and how to sample~$A_i$ given~$A$ and~$u$.}
\end{supplement}

%%  imsart-nameyear.bst  will be used to                   %%
%%  MR numbers will be added by VTeX.                      %%
%%                                                         %%
%%  Use \cite{...} to cite references in text.             %%
%%                                                         %%
%%%%%%%%%%%%%%%%%%%%%%%%%%%%%%%%%%%%%%%%%%%%%%%%%%%%%%%%%%%%%

%%% if your bibliography is in bibtex format, uncomment commands:
%\bibliographystyle{imsart-nameyear} % Style BST file
%\bibliography{bibliography}       % Bibliography file (usually '*.bib')

\clearpage

\begin{frontmatter}
\newcommand{\letitre}{Supplement to ``Set-valued data analysis for interlaboratory comparisons''}
\title{\letitre}
%\title{A sample article title with some additional note\thanksref{t1}}
\runtitle{\letitre}

\begin{aug}
%%%%%%%%%%%%%%%%%%%%%%%%%%%%%%%%%%%%%%%%%%%%%%%
%% Only one address is permitted per author. %%
%% Only division, organization and e-mail is %%
%% included in the address.                  %%
%% Additional information can be included in %%
\author[A]{\fnms{Sébastien J.}~\snm{Petit}},
\author[A]{\fnms{Sébastien}~\snm{Marmin}}
\and
\author[A]{\fnms{Nicolas}~\snm{Fischer}}
%%%%%%%%%%%%%%%%%%%%%%%%%%%%%%%%%%%%%%%%%%%%%%
%% Addresses                                %%
%%%%%%%%%%%%%%%%%%%%%%%%%%%%%%%%%%%%%%%%%%%%%%
\address[A]{Department of Data Science and Uncertainty,
Laboratoire national de métrologie\printead[presep={ ,\ }]{e1,e2,e3}}

\end{aug}

\end{frontmatter}
%%%%%%%%%%%%%%%%%%%%%%%%%%%%%%%%%%%%%%%%%%%%%%
%% Please use \tableofcontents for articles %%
%% with 50 pages and more                   %%
%%%%%%%%%%%%%%%%%%%%%%%%%%%%%%%%%%%%%%%%%%%%%%
%\tableofcontents

\section{Introduction}

This is the Supplementary Material to \citet{petit2025set}.
Details on how to precompute the values of~$D(q, \, i, \, A)$
and how to sample~$A_i$ given~$A$ and~$u$ are presented.

\section{Precomputing the values of~$D(q, \, i, \, A)$}

To compute the marginal likelihoods
$$P ( X_i^{(1)}, \dots, X_i^{(p_i)} \, \lvert \, A, \, u ),$$
the values
$$
D(q, \, i, \, A) =
\sum_{\underline{r} = (r_1, \dots, r_{p_i}) \in \{ 0, \dots, n \}^{p_i}}
\mathcal{C}_n \left( A, \, \underline{X}_i; \, q, \, \underline{r} \right)
\frac{ \int \prod_{j=1}^{p_i} e_{u_i} \left( n - r_j \right) g(u_i) \mathrm{d}u_i}{
\prod_{j=1}^{p_i} \mathcal{C}_n \left( r_j \right)
},
$$
introduced by \citet[Section 5.2]{petit2025set} must be computed efficiently
for all~$0 \leq q \leq n$, all~$A \in \Pn$,
and all~$1 \leq i \leq L$.

The integrals~$\int \prod_{j=1}^{p_i} e_{u_i} \left( n - r_j \right) g(u_i) \mathrm{d}u_i$
can be computed either analytically or numerically. Computation time is not an issue, since
they can be precomputed once and for all and then accessed from a hash table indexed
by the tuple~$(r_1, \dots, r_{p_i})$.
To compute~$D(q, \, i, \, A)$ for~$i$,~$A$, and~$q$,
the main bottleneck is
computing~$\mathcal{C}_n \left( A, \, \underline{X}_i; \, q, \, \underline{r} \right)$,
for all~$\underline{r}$.

Let~$p \geq 1$ and~$\underline{Y} = (Y_1, \dots, Y_p) \in \Pn^p$ be arbitrary.
For $I \subset \{ 1, \dots, p \}$, write
\begin{equation*}
\alpha_{I}(\underline{Y}) =  \left( \Ocal \cap \left( \cap_{j \in I} Y_j \right) \right) \setminus \left( \cup_{j \notin I} Y_j \right).
\end{equation*}
The sets~$\alpha_I(\underline{Y})$, for~$I \subset \{ 1, \dots, p \}$, partition~$\Ocal$ with respect to membership in the~$Y_j$s
and we have
\begin{equation}\label{eq:expr_Cn}
\mathcal{C}_n \left( \underline{Y}; \, \underline{r} \right)
=
\sum_{s \in \Scal(\underline{Y}; \underline{r})} \ \prod_{I \subset \{ 1, \dots, p \}} \binom{\# \alpha_I(\underline{Y})}{s_I},
\end{equation}
with
$$
\Scal(\underline{Y}; \underline{r}) =  \left\{ 
s \in \Scal(\underline{Y}) \, \colon \quad
\sum_{j \in I \subset \{ 1, \dots, p \}} s_I = r_j, \quad \forall 1 \leq j \leq p \,
\right\}
$$
and
$$
\Scal(\underline{Y}) =  \left\{ 
s \in \prod_{I \subset \{ 1, \dots, p \}} \{ 0, \dots, \# \alpha_{I}(\underline{Y}) \} \, \colon \,
\sum_{I \subset \{ 1, \dots, p \}} s_I = n 
\right\}.
$$
The elements of~$\Scal(\underline{Y}; \underline{r})$ represent the possible counts~$\# \left( Y \cap \alpha_I(\underline{Y}) \right)$,
for a~$Y \in \Pn$ such that~$\# (Y \cap Y_j) = r_j$.

The elements of~$\Scal(\underline{Y})$ are the valid solutions
to the problem of selecting~$n$ items in~$2^{p}$ urns and
can be generated using recursive programming.
To compute~\eqref{eq:expr_Cn}
for all~$\underline{r}$ we start by generating the elements
of~$\Scal(\underline{Y})$ and partition them
using a hash table indexed by the
tuples~$(\sum_{I \ni 1} s_I, \dots, \sum_{I \ni p} s_I)$.
Then, after this preprocessing step, the value~\eqref{eq:expr_Cn}
can be computed easily for all~$\underline{r}$. 

To handle the EMC interlaboratory comparison,
the count~$\mathcal{C}_n \left( A, \, \underline{X}_i; \, q, \, \underline{r} \right)$
must be computed for each~$i$,~$q \in \{0, \dots, n\}$,~$\underline{r} = (r_1, \dots, r_{p_i}) \in \{ 0, \dots, n \}^{p_i}$,
and~$A \in \Pn$.
For a brief moment, we will omit the dependence on~$i$ and write~$p$ and~$\underline{X}$ to simplify the notations.
Computations can be saved by observing
that~$\mathcal{C}_n \left( A, \, \underline{X}; \, q, \, \underline{r} \right)$ depends on~$A$
only through the counts $\# (A \cap \alpha_{I}(\underline{X}))$, for~$I \subset \{ 1, \dots, p \}$.
Indeed,
observe first in~\eqref{eq:expr_Cn}
that~$\mathcal{C}_n \left( A, \, \underline{X}; \, q, \, \underline{r} \right)$
depends on~$A$ only through the counts~$\# \alpha_{I}(\underline{X}, A)$, $I \subset \{ 1, \dots, p + 1\}$.
Then, one has:
\begin{equation}\label{eq:rec_alpha}
  \left\{
    \begin{array}{lcl}
	  \alpha_{I}(\underline{X}, A) & = & \alpha_{I}(\underline{X}) \setminus A, \\[5pt]
      \alpha_{I \cup \{p + 1\}}(\underline{X}, A) & = & \alpha_{I}(\underline{X}) \cap A,
    \end{array}
  \right.
\quad
\mathrm{for} \ \
I \subset \{ 1, \dots, p\}.
\end{equation}
From this, it can be seen that, given~$\underline{X}$,
the counts~$\# \alpha_{I}(\underline{X}, A)$, $I \subset \{ 1, \dots, p + 1\}$, depends
only on the counts~$\# (\alpha_{I}(\underline{X}) \cap A)$, $I \subset \{ 1, \dots, p\}$.
The set of all possible values of these counts over all~$A \in \Pn$
is~$\Scal(\underline{X})$.

Unfortunately, the previous approach does not make it possible
to handle the EMC interlaboratory comparison
with a reasonable computation time,
but the following tricks do.
Let~$\underline{r} \in \{ 0, \dots, n \}^{p}$,~$s \in \Scal(\underline{X}; \underline{r})$, and~$A \in \Pn$.
Write~$\mathbb{N}_0$ for the set of non-negative integers and~$\mathcal{P}(\{1, \dots, p + 1\})$
for the power set of~$\{1, \dots, p + 1\}$.
The set
\begin{equation*}
T(s) =
\left\{  \tilde{s} \in \mathbb{N}_0^{\mathcal{P}(\{1, \dots, p + 1\})}
\quad
\colon
\quad
\tilde{s}_I + \tilde{s}_{I \cup \{p + 1 \}} = s_I,
\quad
\mathrm{for} \ \
I \subset \{ 1, \dots, p\}
\right\}
\end{equation*}
can be generated from~$s$ using recursive programming and then be split into
the sets
$$T(s; q) = \left\lbrace \tilde{s} \in T(s) \, \colon \, \sum_{I \subset \{ 1, \dots, p\}} \tilde{s}_{I \cup \{p + 1 \}} = q \right\rbrace,$$
for~$0 \leq q \leq n$.
For~$A \in \Pn$,
it can be checked that
\begin{equation*}
\Scal(\underline{X}, A;  \underline{r}, q) = \left\{  \tilde{s} \in \bigcup_{s \in \Scal(\underline{X}; \underline{r})} T(s; \, q) \ \colon \ \tilde{s}_I \leq \# \alpha_{I}(\underline{X}, A), \quad \mathrm{for} \ I \subset \{ 1, \dots, p+ 1\} \right\}.
\end{equation*}
Algorithm~\ref{alg:brute_force-effet_op-inference_scheme}
builds on this equality to calculate efficiently the quantities~$D(q, i, A)$.

\begin{algorithm}
\caption{Algorithm for precomputing the values of~$D(q, A)$ (the subscripts~$i$ have been removed for readability).
Recall from~\eqref{eq:expr_Cn} and~\eqref{eq:rec_alpha} that~$D(q, A)$ depends on~$A$ only through
the counts~$\# (\alpha_{I}(\underline{X}) \cap A)$, $I \subset \{ 1, \dots, p\}$.
Consequently, we will write~$D(q, \overline{s})$, with~\hbox{$\overline{s} \in \Scal(\underline{X})$}.
We also use the convention that the value of a hash table is zero if it does not contain the key.}\label{alg:brute_force-effet_op-inference_scheme}
\begin{algorithmic}
\vspace{0.2cm}
\State Initialize an empty hash table~$D$
\vspace{0.2cm}
\State Build the sets~$\Scal(\underline{X}; \underline{r})$, for~$\underline{r} \in \{ 0, \dots, n \}^{p}$, using recursive programming
\vspace{0.2cm}
\For{$\underline{r} = (r_1, \dots, r_p) \in \{ 0, \dots, n \}^{p}$}
\vspace{0.2cm}
\State
Compute
$G = \frac{ \int \prod_{j=1}^{p} e_{u} \left( n - r_j \right) g(u) \mathrm{d}u}{
\prod_{j=1}^{p} \mathcal{C}_n \left( r_j \right)}$
\vspace{0.2cm}
\For{$s \in \Scal(\underline{X}; \underline{r})$}
\vspace{0.2cm}
\State Generate the sets~$T(s; q)$, for~$0 \leq q \leq n$, using recursive programming
\vspace{0.2cm}
\EndFor
\vspace{0.2cm}
\For{$q \in \{ 0, \dots, n \}$}
\vspace{0.2cm}
\For{$\tilde{s} \in \bigcup_{s \in \Scal(\underline{X}; \underline{r})} T(s; \, q)$}
\vspace{0.2cm}
\For{$\overline{s} \in \Scal(\underline{X})$} \ (This represents the possible values of Equation~\ref{eq:rec_alpha} over~$A \in \Pn$)
\vspace{0.2cm}
\If{$\tilde{s}_{I \cup {p + 1}} \leq \overline{s}_I$ and~$\tilde{s}_{I} \leq \# \alpha_I(\underline{X}) - \overline{s}_I$, for all~$I \subset \{ 1, \dots, p\}$,}
\vspace{0.2cm}
\State $D(q, \, \overline{s}) = D(q, \, \overline{s}) 
+ G \cdot \prod_{I \subset \{ 1, \dots, p \}}
\binom{\# \alpha_I(\underline{X}) - \overline{s}_I}{\tilde{s}_I}
\binom{\overline{s}_I}{\tilde{s}_{I \cup {p + 1}}}
$
\vspace{0.2cm}
\EndIf
\EndFor
\EndFor
\EndFor
\EndFor
\end{algorithmic}
\end{algorithm}

\section{Sampling $A_i$ given~$A$ and~$u$}

Use again the notations $\underline{X}_i$ and $p_i$ for the $i$-th laboratory.
The preceding tools make it possible to sample~$A_i$ given~$A$ and~$u$.
Consider the second equality for~$P ( X_i^{(1)}, \dots, X_i^{(p_i)} \, \lvert \, A, \, u  )$
given by \citet[Section~5.2]{petit2025set}.
One can first sample~$(\underline{r}, \, q) \in \{ 0, \dots, n \}^{p_i + 1}$ according
to the unnormalized probability mass function
$$
\mathcal{C}_n \left( \underline{X}_i, \, A; \, \underline{r}, \, q \right)
\frac{
e_u \left( n - q \right)
\int \prod_{j = 1}^{p_i} e_{u_i} \left( n - r_j \right) g(u_i) \mathrm{d}u_i
}{
\mathcal{C}_n \left( q \right)
\prod_{j = 1}^{p_i}
\mathcal{C}_n \left( r_j \right)
}.
$$
Then, given~$(\underline{r}, \, q)$, one can sample~$s \in \Scal(\underline{X}_i, A; \underline{r}, q)$
according to the (unnormalized) probability mass function
$$
\prod_{I \subset \{ 1, \dots, p_i + 1 \}} \binom{\# \alpha_I(\underline{X}_i, A)}{s_I}
$$
given by~\eqref{eq:expr_Cn}.
Finally, one obtains a sample from~$A_i$ given~$A$ and~$u$ by sampling
uniformly without replacement~$s_I$ elements from~$\alpha_I(\underline{X}_i, A)$,
for each~$I \subset \{ 1, \dots, p_i + 1 \}$.

%% if your bibliography is in bibtex format, uncomment commands:
\bibliographystyle{imsart-nameyear} % Style BST file
\bibliography{bibliography}       % Bibliography file (usually '*.bib')

\end{document}